# Enhancement of Noisy Speech with Low Speech Distortion Based on Probabilistic Geometric Spectral Subtraction

Md Tauhidul Islam[a], Celia Shahnaz[b,*], Wei-Ping Zhu[c], M. Omair Ahmad[c]

[a]*Department of Electrical and Computer Engineering, Texas A&M University, College Station, Texas, USA-77840*
[b]*Department of Electrical and Electronic Engineering, Bangladesh University of Engineering and Technology, Dhaka-1000, Bangladesh*
[c]*Department of Electrical and Computer Engineering, Concordia University, Montreal, Quebec H3G 1M8, Canada*

**Abstract**

A speech enhancement method based on probabilistic geometric approach to spectral subtraction (PGA) performed on short time magnitude spectrum is presented in this paper. A confidence parameter of noise estimation is introduced in the gain function of the proposed method to prevent subtraction of the overestimated and underestimated noise, which not only removes the noise efficiently but also prevents the speech distortion. The noise compensated magnitude spectrum is then recombined with the unchanged phase spectrum to produce a modified complex spectrum prior to synthesize an enhanced frame. Extensive simulations are carried out using the speech files available in the NOIZEUS database in order to evaluate the performance of the proposed method.

## 1. Introduction

Speech enhancement methods are important in practical noisy situations, where intelligibility and quality of speech are often degraded by various background noise sources. These methods can be mainly classified in three categories based on their domains of operation, namely time-domain, frequency-domain and time-frequency domain. Subspace based approach is the prominent time domain based method [1, 2, 3]. A large computational load is the main problem of this approach. In frequency-domain based methods, such as spectral subtraction [4, 5, 6, 7, 8], mean square error estimator (MMSE) [9, 10], Wiener filter [11] and Kalman fillter [12], this problem is solved. But an estimation of noise spectrum is necessary in these methods. Another problem with the MMSE and Weiner filtering methods is that they cannot control the trade-off between the preservation of speech harmonics and noise removal. Time-frequency domain methods involve the employment of the family of wavelets or wavelet packets [13, 14, 15, 16]. The wavelet or wavelet packet based methods are computationally slow like time-domain methods.

Low computational complexity with better noise suppression ability is the main charm of the spectral subtraction methods [4, 17, 18]. Most of the methods based on spectral subtraction assume that the noise and speech are uncorrelated in the short frame it is applied, which is not true for most of the practical scenarios. Another problem of the spectral subtraction is the musical noise, which is sometimes more disturbing than the original noise. Musical noise arises because of putting negative values to zero obtained for some points, where noise spectrum has higher values than the speech spectrum. This noise arises in time domain as random from frame to frame. A geometric approach of spectral subtraction (GA) is proposed in [6] as a solution to these problems. In this method, the speech signal and noise are taken as two vectors and a gain function is derived without taking their cross-correlation as zero. The main assumption of this method is that the noise estimation is accurate, which is not true for most of the practical cases.

In this paper, we propose a probabilistic modification of GA, where a confidence parameter of noise estimation is introduced in the gain function of GA to prevent the speech distortion and noise residual in the enhanced speech.

The paper is organized as follows. Section 2 presents the problem formulation and proposed method. Section 3 describes results. Concluding remarks are presented in Section 4.

---

*Corresponding author
 Email address:* `celia.shahnaz@gmail.com` (Celia Shahnaz)

## 2. Problem Formulation and Proposed Method

In the presence of additive noise denoted as d[n], a clean speech signal x[n] gets contaminated and produces noisy speech y[n]. The proposed method is based on the AMS framework, where speech is analyzed frame wise since it can be assumed to be quasi-stationary. The noisy speech is segmented into overlapping frames by using a sliding window. A windowed noisy speech frame can be expressed in the time domain as

$$y^\tau[n] = x^\tau[n] + d^\tau[n], \tag{1}$$

where $\tau$ is the frame number, $\tau = 1, \ldots, T$, $T$ is the total number of frame. If $Y^\tau[k]$, $X^\tau[k]$ and $D^\tau[k]$ are the fast Fourier transform (FFT) representations of $y^\tau[n]$, $x^\tau[n]$ and $d^\tau[n]$, respectively, we can write

$$Y^\tau[k] = X^\tau[k] + D^\tau[k], \tag{2}$$

where $k = 0, 1, 2, \ldots N-1$, $N$ is the length of a frame in samples. The $N$-point FFT, $Y^\tau[k]$ of $y[n]$ can be computed as

$$Y^\tau[k] = \sum_{n=0}^{N-1} y^\tau[n] e^{-\frac{j2\pi nk}{N}}. \tag{3}$$

$Y^\tau[k]$ is modified in the proposed method to obtain an estimate of $X^\tau[k]$. An overview of the proposed speech enhancement method is shown by a block diagram in Fig. 1. It is seen from Fig. 1 that Fourier transform is first applied to each input speech frame. A confidence parameter of noise estimation is determined based on the *a posteriori* and *a priori* SNRs of the noisy speech spectrum. The magnitude of the noisy speech spectrum is compensated based on this confidence parameter. The modified magnitude with unchanged phases is used to obtain a modified complex spectrum. An enhanced speech frame is obtained via inverse Fourier transform of the modified complex spectrum and overlap and add method.

### 2.1. Probabilistic geometric approach

From the basic rule of spectral subtraction, $|X^\tau[k]|$ can be written as

$$|\widehat{X}^\tau[k]|^2 = |Y^\tau[k]|^2 - |\widehat{D}^\tau[k]|^2, \tag{4}$$

which can also be written in the following form,

$$|\widehat{X}^\tau[k]|^2 = |H^\tau_{PGA}[k]|^2 |Y^\tau[k]|^2, \tag{5}$$

where $H^\tau_{PGA}[k]$ is the gain function of PGA for $\tau^{\text{th}}$ frame. Geometrically, $Y^\tau[k]$ can be represented as sum of two complex number $X^\tau[k]$ and $D^\tau[k]$ which is shown in Fig. 2. $Y^\tau[k]$, $X^\tau[k]$ and $D^\tau[k]$ are the complex numbers which can be expressed in polar form as

$$Y^\tau[k] = a_Y e^{j\theta_Y}, X^\tau[k] = a_X e^{j\theta_X}, D^\tau[k] = \rho a_D e^{j\theta_D}, \tag{6}$$

where $\rho$ is a constant dependent on the *a posteriori* and *a priori* SNRs. We drop the superscript $\tau$ and index $k$ for $\rho$ for reducing the notational complexity in the next derivation. We call $\rho$ the confidence parameter of noise estimation as it considers the signal and noise power in a frame of noisy speech. We will discuss detail on determination of $\rho$ in subsection 2.1.3. Substituting the values of $Y^\tau[k]$, $X^\tau[k]$ and $D^\tau[k]$ in (2), we obtain

$$a_Y e^{j\theta_Y} = a_X e^{j\theta_X} + \rho a_D e^{j\theta_D}, \tag{7}$$

where $[a_Y, a_X, a_D]$ are the magnitudes and $\theta_Y, \theta_X, \theta_D$ are the phases of noisy, clean and noise spectra respectively. If we express these complex numbers in a right angle triangle ABC as shown in Fig. 3, using sine rule, we can write

$$\overline{AB} = a_Y \sin(\theta_D - \theta_Y) = a_X \sin(\theta_D - \theta_X) \tag{8}$$

$$\Rightarrow a_Y^2 \sin^2(\theta_D - \theta_Y) = a_X^2 \sin^2(\theta_D - \theta_X) \tag{9}$$

$$\Rightarrow a_Y^2 [1 - \cos^2(\theta_D - \theta_Y)] = a_X^2 [1 - \cos^2(\theta_D - \theta_X)] \tag{10}$$

$$\Rightarrow a_Y^2 [1 - C_{YD}^2] = a_X^2 [1 - C_{XD}^2]. \tag{11}$$

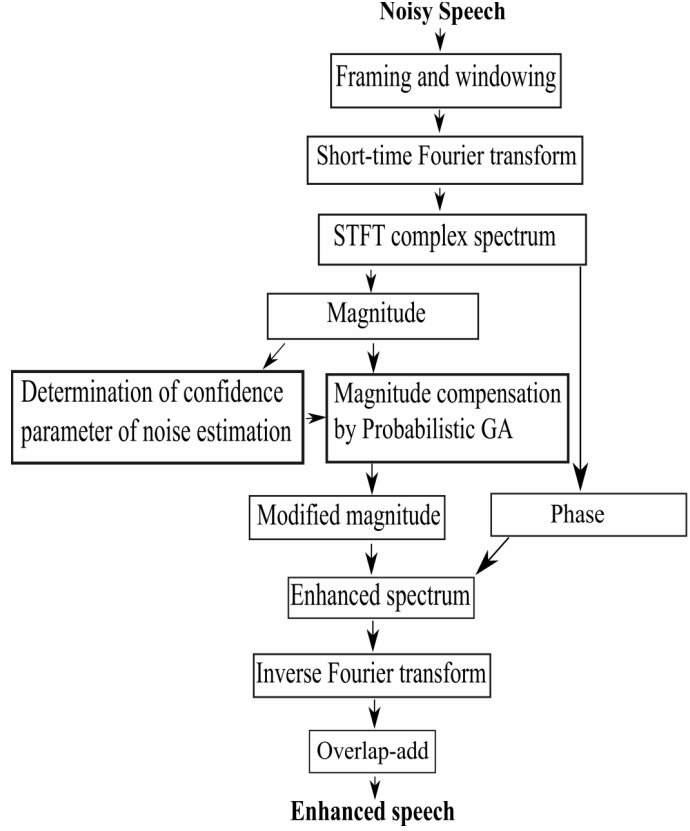

Figure 1: Block diagram representation of the proposed method.

The gain function $H_{PGA}$ can be defined as

$$H^\tau_{PGA}[k] = \sqrt{\frac{a_Y^2}{a_X^2}} = \sqrt{\frac{1-c_{YD}^2}{1-c_{XD}^2}}, \qquad (12)$$

where $c_{YD} = cos(\theta_D - \theta_Y)$ and $c_{XD} = cos(\theta_D - \theta_X)$.

When $X$ and $D$ are orthogonal, $c_{XD}$ becomes zero. It is the case when noise and signal are uncorrelated and the cross-terms between $X$ and $D$ are zero. Using cosine rules in triangle ABC, we can write

$$c_{YD} = \frac{a_Y^2 + \rho^2 a_D^2 - a_X^2}{2 a_Y \rho a_D}, \qquad (13)$$

$$c_{XD} = \frac{a_Y^2 - \rho^2 a_D^2 - a_X^2}{2 a_X \rho a_D}. \qquad (14)$$

If we divide the equation by $a_D^2$ we get

$$c_{YD} = \frac{\frac{a_Y^2}{a_D^2} + \rho^2 - \frac{a_X^2}{a_D^2}}{\frac{2a_Y}{a_D}\rho} = \frac{\gamma + \rho^2 - \xi}{2\sqrt{\gamma}\rho}, \qquad (15)$$

$$c_{YD} = \frac{\frac{a_Y^2}{a_D^2} - \rho^2 - \frac{a_X^2}{a_Y^2}}{\frac{2a_Y}{a_D}\rho} = \frac{\gamma - \rho^2 - \xi}{2\sqrt{\xi}\rho}. \qquad (16)$$

The variables $\gamma$ and $\xi$ are defined as follows:

$$\gamma = \frac{a_Y^2}{a_D^2}, \quad \xi = \frac{a_X^2}{a_D^2}. \tag{17}$$

As defined in [9], $\gamma$ and $\xi$ are the *a posterior* and *a priori* SNRs, respectively used in MMSE algorithm. Substituting values of $\gamma$ and $\xi$ in (12), we obtain the expression of $H_{PGA}$ as

$$H_{PGA}^{\tau}[k] = \sqrt{\frac{1 - (\frac{\gamma + \rho^2 - \xi}{2\sqrt{\gamma}\rho})^2}{1 - (\frac{\gamma - \rho^2 - \xi}{2\sqrt{\xi}\rho})^2}}. \tag{18}$$

The equation of $H_{PGA}$ is different from the gain function of GA defined in [6], as it has a confidence parameter of noise estimation, $\rho$ in the expression. When $\rho = 1$, the gain function of GA and PGA become equal, i.e., $|H_{PGA}| = |H_{GA}|$. We will discuss the impact of $\rho$ on $H_{PGA}$ for two different cases, i.e., overestimated noise and underestimated noise.

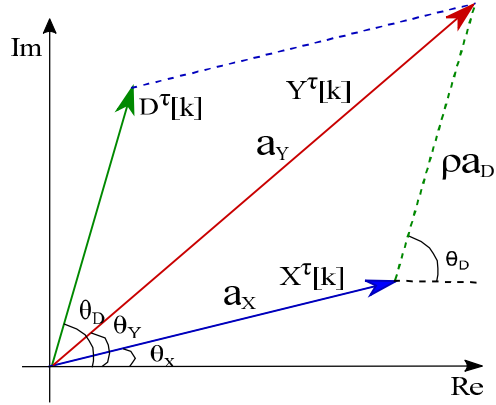

Figure 2: Representation of the noisy speech spectrum Y[k] in the complex plane as the sum of the clean speech spectrum X[k] and noise spectrum D[k].

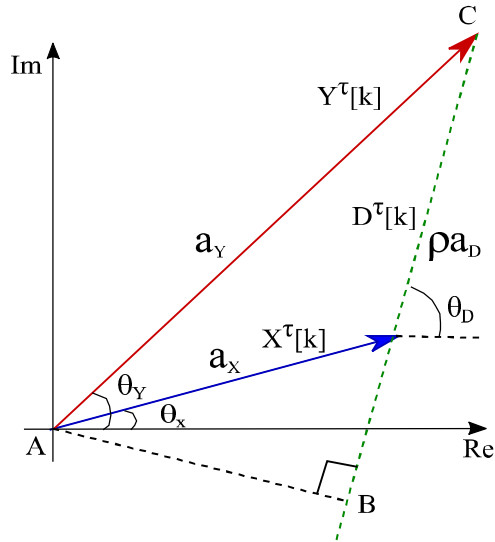

Figure 3: Triangle representation of the phases of the noisy speech spectrum Y[k], clean speech spectrum X[k] and noise spectrum D[k].

### 2.1.1. Effect of $\rho$ on $H_{PGA}$ for overestimated noise

We can define a case of noise overestimation through the *a posteriori* SNR $\gamma$ and the *a priori* SNR $\xi$. If $\xi$ is significantly higher than $\gamma$, it can be assumed that too much noise is deducted from the noisy speech spectrum, which implies that noise is overestimated. For such a case, when $\gamma = 1$ and $\xi = 10$, we plot $H_{PGA}$ for different values of $\rho$ in Fig. 4. From this figure, we see that with decrement of $\rho$, $H_{PGA}$ increases. As $\rho$ denotes the confidence parameter of noise estimation, decrement of $\rho$ implies worse noise estimation. Gain function is increased in this case so that less subtraction is imposed on the noisy speech spectrum.

### 2.1.2. Effect of $\rho$ on $H_{PGA}$ for underestimated noise

If $\gamma$ is significantly larger than $\xi$, it can be said that too much noise is still in the denoised signal, which implies that the noise is underestimated. For such a case, when $\gamma = 10$ and $\xi = 1$, we plot $H_{PGA}$ for different values of $\rho$ in Fig. 5. From this figure, we see that with decrement of $\rho$, $H_{PGA}$ decreases to impose more noise reduction on the noisy signal spectrum and thus ensure removal of the residual noise that was left because of underestimation of noise.

Plots of the gain function of GA, $H_{GA}$ and of PGA, $H_{PGA}$ are shown in Fig. 6 for the spectrum of a noisy speech frame from NOIZEUS database. From this figure, we see that $H_{PGA}$ has different values than $H_{GA}$ at some points, where $\rho$ is less than 1. These points are the points, where we overestimated or underestimated noise. Thus, in PGA, by increasing or decreasing $H_{PGA}$, the speech distortion or noise residual is prevented.

### 2.1.3. Determination of $\rho$

We propose to determine $\rho$ at $\tau^{th}$ frame and $k^{th}$ sample, $\rho^\tau[k]$ by

$$\rho^\tau[k] = \sqrt{1 - P^\tau_{local}[k] P^\tau_{global}[k] P_{frame}(\tau)}, \tag{19}$$

where $P^\tau_{local}[k]$ and $P^\tau_{global}[k]$ are determined from the following equation,

$$P^\tau_\psi[k] = \begin{cases} 0, & \text{if } \xi^\tau_\psi[k] \leq \xi_{min}, \\ 1, & \xi^\tau_\psi[k] \geq \xi_{max}, \\ \frac{\log(\xi^\tau_\psi[k]/\xi_{min})}{\log(\xi_{max}/\xi_{min})}, & \text{otherwise,} \end{cases} \tag{20}$$

where the subscript $\psi$ denotes either "local" or "global" and $\xi^\tau_\psi[k]$ represents either "local" or "global" mean values of the *apriori* SNR [19, 9]. $\xi^\tau_\psi[k]$ is defined as

$$\xi^\tau_\psi[k] = \sum_{i=-W_\psi}^{i=W_\psi} h_\psi(i) \xi^\tau[k-i], \tag{21}$$

where $h_\psi$ is a window function of length of $2W_\psi + 1$. In this equation, $\xi^\tau[k]$ is defined as

$$\xi^\tau[k] = (1 - \alpha_\xi)\hat{\gamma}^\tau[k], \tag{22}$$

where $\alpha_\xi$ is a constant and $\hat{\gamma}$ is the estimated *a posteriori* SNR.

In (19), $P_{frame}(\tau)$ is determined as

$$P_{frame}(\tau) = \begin{cases} 0, & \text{if } \xi_{frame}(\tau) < \xi_{min}, \\ 1, & \text{if } \xi_{frame}(\tau) > \xi_{frame}(\tau - 1) \text{ and} \\ & \xi_{frame}(\tau) > \xi_{min}, \\ \mu(\tau), & \text{otherwise,} \end{cases} \tag{23}$$

where $\mu(\tau)$ is determined by

$$\mu(\tau) = \begin{cases} 0, & \text{if } \xi_{frame}(\tau) \leq \xi_{peak}\xi_{min} \\ 1, & \text{if } \xi_{frame}(\tau) \geq \xi_{peak}\xi_{max}, \\ \frac{\log(\xi_{frame}(\tau)/(\xi_{peak}\xi_{min}))}{\log(\xi_{max}/\xi_{min})}, & \text{otherwise.} \end{cases} \tag{24}$$

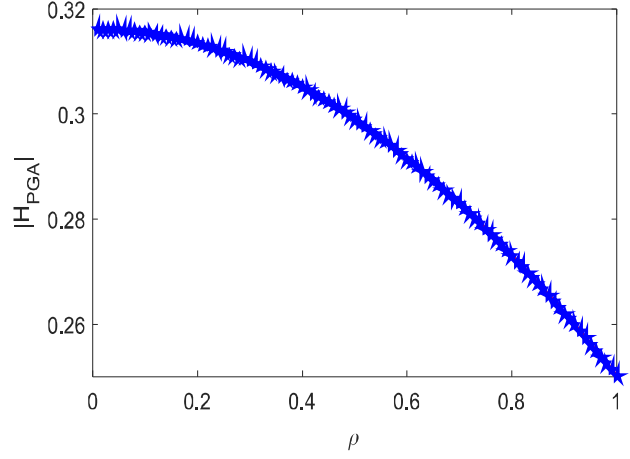

Figure 4: Plot of gain function of PGA for different values of $\rho$ for overestimated noise.

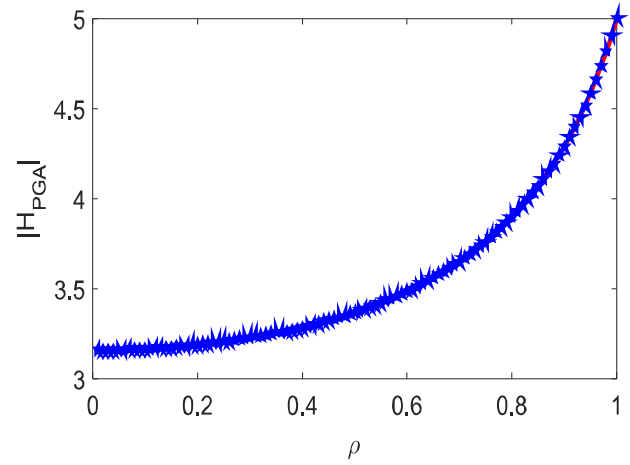

Figure 5: Plot of gain function of PGA for different values of $\rho$ for underestimated noise.

In (20), (23) and (24), $\xi_{min}$, $\xi_{max}$ and $\xi_{peak}$ are empirically determined constants and $\xi_{frame}(\tau)$ is defined as

$$\xi_{frame}(\tau) = \frac{1}{N} \sum_{k=0}^{N-1} \xi^{\tau}[k]. \qquad (25)$$

*2.2. Resynthesis of enhanced signal*

In the synthesis stage, the unchanged phase spectrum is recombined with the compensated magnitude spectrum in order to produce an enhanced complex spectrum.

$$\widehat{X^{\tau}}[k] = |\widehat{X^{\tau}}[k]|e^{\angle Y^{\tau}[k]}. \qquad (26)$$

The enhanced speech frame is synthesized by performing the inverse FFT on the resulting complex spectrum as

$$\widehat{x^{\tau}}[n] = Re\left(IFFT\{\widehat{X^{\tau}}[k]\}\right), \qquad (27)$$

where $Re(\cdot)$ denotes the real part of the number inside it and $\widehat{x}[n]$ represents the the enhanced speech frame. The final enhanced speech signal is synthesized by using the standard overlap and add method [20].

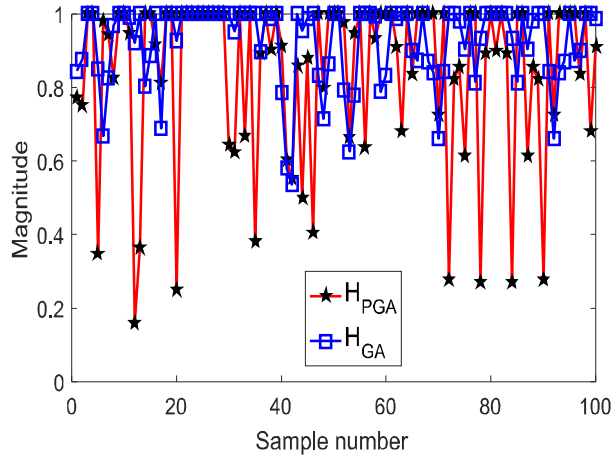

Figure 6: Gain functions of the geometric approach and proposed method.

## 3. Results

In this Section, a number of simulations is carried out to evaluate the performance of the proposed method.

*3.1. Implementation*

The proposed method is implemented in MATLAB R2016b graphical user interface development environment (GUIDE). The MATLAB software with its user manual is attached as supplementary material with the paper. This software also includes the implementation of a classical and a recent speech enhancement methods, i.e., GA [6] and universal thresholding method (UTM) [13]. GA's code is taken from http://ecs.utdallas.edu/loizou/cimplants/ and MATLAB implementation of UTM is used. The MATLAB implementations of the calculation of segmental and overall SNR improvement are taken from http://ecs.utdallas.edu/loizou/cimplants/ [21]. The noise is estimated from the initial silence frames in the proposed method. The noise is estimated in GA by using the method described in [22] and the MATLAB code for this method has been acquired from https://www.uni-oldenburg.de/en/mediphysics-acoustics/spe-echsignal-processing/publications/noise-power-estimation/.

*3.2. Simulation Conditions*

Real speech sentences from the NOIZEUS database are used for the simulations, where the speech data is sampled at 8 kHz. To imitate a practical noisy environment, noise sequence is added to the clean speech samples at different SNR levels ranging from 10 dB to −20 dB. Two different types of noise, such as, airport and babble are adopted from the NOIZEUS database.

In order to obtain overlapping analysis frames, hamming windowing operation is performed, where the size of each frame is 100 samples with 50% overlap between successive frames. The values of the constants used to determine the confidence parameter of noise estimation are given in Table 1.

*3.3. Comparison Metrics*

Standard Objective metrics [21], namely, segmental SNR (SNRSeg) improvement in dB, overall SNR improvement in dB and perceptual evaluation of speech quality (PESQ) are used for the evaluation of PGA method. The proposed PGA method is subjectively evaluated in terms of the spectrogram representations of the clean speech, noisy speech and enhanced speech. Formal listening tests are also carried out in order to find the analogy between the objective metrics and subjective sound quality. The performance of PGA is compared with GA [6] and UTM [13] in both objective and subjective senses.

Table 1: Constants used to determine the confidence parameter of noise estimation

| Constants | Value of constants |
|---|---|
| $\xi_{min}$ | -10 dB |
| $\xi_{max}$ | -5 dB |
| $\xi_{peak}$ | 10 dB |
| $w_{local}$ | 1 |
| $w_{global}$ | 15 |
| $\alpha_\xi$ | 0.7 |

*3.4. Objective Evaluation*

Two commonly used noise types, airport and multi-talker babble are chosen for the objective evaluation.

*3.4.1. Results for speech signals with airport noise*

SNRSeg improvement, overall SNR improvement and PESQ scores for speech signals corrupted with airport noise for GA, UTM and PGA are shown in Fig. 7, 8 and Table 2.

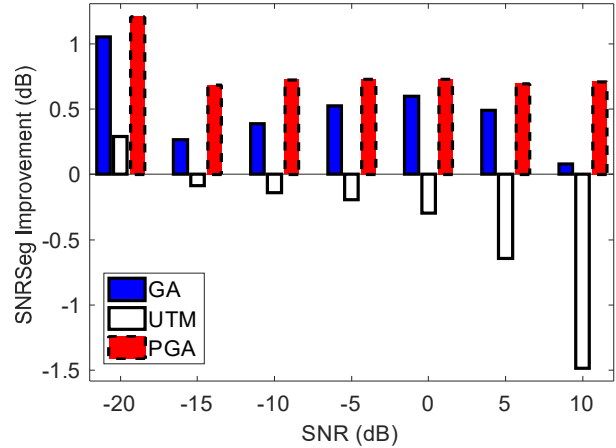

Figure 7: SNRSeg improvement for different methods in airport noise.

In Fig. 7, the performance of the proposed method is compared with those of the other methods at different levels of SNR for airport noise in terms of SNRSeg improvement. We see from this figure that GA and PGA produces enhanced speeches with positive SNRSeg improvement for the whole SNR levels of −20 to 10 dB, whereas UTM provides negative SNRSeg improvement at −15 to 10 dB. PGA shows the highest SNRSeg improvement of 1.4 dB at −20 dB, which is significantly higher than other two methods at same SNR. This trend continues for almost the whole SNR range, which proves the efficacy of PGA to produce better enhanced speech.

Fig. 8 represents the overall SNR improvements in dB as a function of SNR for PGA, GA and UTM. It is clearly seen from this figure that PGA provides higher overall SNR improvements at all the SNR levels from −20 to 10 dB in comparison to other two methods. At a lower SNR as −20 dB, PGA provides overall SNR improvement of 9 dB, which is quite better than GA and UTM.

PESQ scores for UTM, GA and PGA are shown in Table 2 for SNR range of −20 to 10dB. We realize from this table that for most of the SNR levels, PGA shows better PESQs in comparison to UTM and GA. PGA provides PESQ value of 2.15 at 10 dB, which is significantly higher than GA and UTM. As a higher PESQ score indicates a better speech quality at a particular SNR, PGA is indeed better in performance in presence of airport noise.

Table 2: PESQ for different methods in airport noise

| SNR(dB) | UTM  | GA   | PGA  |
|---------|------|------|------|
| -20     | 0.02 | 0.24 | 0.21 |
| -15     | 0.20 | 0.43 | 0.38 |
| -10     | 0.31 | 0.85 | 0.89 |
| -5      | 0.72 | 1.21 | 1.51 |
| 0       | 1.14 | 1.37 | 1.74 |
| 5       | 1.19 | 1.52 | 1.86 |
| 10      | 1.24 | 1.81 | 2.15 |

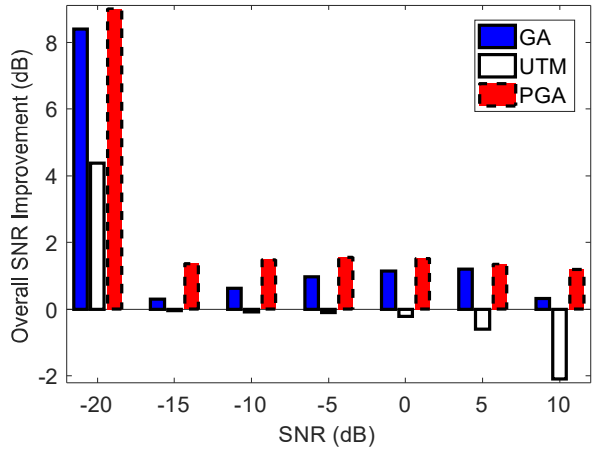

Figure 8: Overall SNR improvement for different methods in airport noise.

*3.4.2. Results for speech signals with multi-talker babble noise*

SNRSeg improvement, overall SNR improvement and PESQ scores for speech signals corrupted with babble noise for GA, UTM and proposed PGA method are shown in Figs. 9, 10 and 11.

In Fig. 9, the performance of the proposed method in the context of SNRSeg improvement is compared with those of the other methods at different levels of SNR. We see that PGA does extremely good in lower SNR levels as −20, −10 or −5 dB. But at higher SNR levels, i.e., 0 and 5 dB, GA provides comparable values to PGA. UTM fails to show any competitive value for the whole SNR range.

Fig. 10 represents the overall SNR improvement as a function of SNR for all the methods. The proposed PGA method shows better performance in comparison to other two methods for almost whole SNR range.

In Fig. 11, It can be seen that at a high level of SNR, such as 5 or 10 dB, all the methods show lower values of PESQ scores than PGA. PGA also yields larger PESQ scores compared to that of the other methods at lower levels of SNR. As a higher PESQ score indicates a better speech quality, PGA is indeed better in performance in comparison to GA and UTM even in presence of babble noise.

*3.5. Subjective Evaluation*

To evaluate the performance of the proposed method and other competing methods subjectively, we use two commonly used tools. The first one is the plot of the spectrograms of the output for all the methods and compare their performance in terms of preservation of harmonics and capability to remove noise.

The spectrograms of the clean speech, the noisy speech, and the enhanced speech signals obtained by using the proposed method and all other methods are presented in Fig. 12 for airport noise-corrupted speech at an SNR of 10 dB.

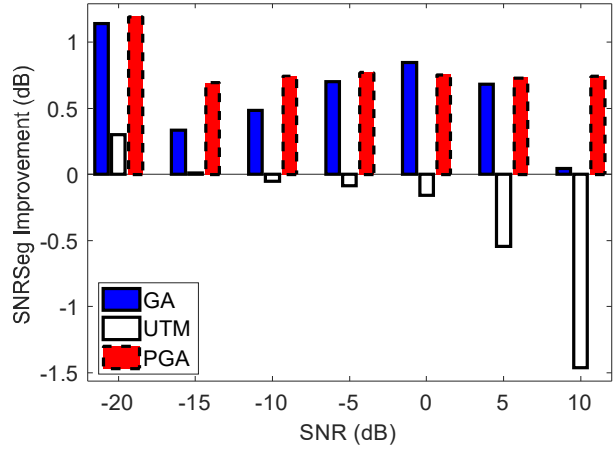

Figure 9: segmental SNR improvement for different methods in babble noise.

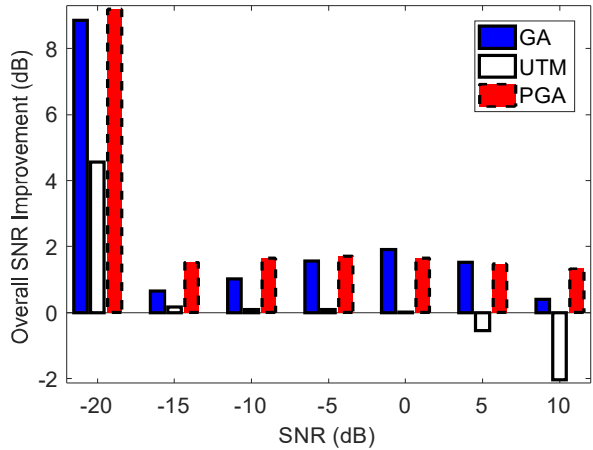

Figure 10: Overall SNR improvement for different methods in babble noise.

It is obvious from the spectrograms that the proposed method preserves the harmonics significantly better than all the other competing methods. The noise is also reduced at every time point for the proposed method which attests our claim of better performance in terms of higher SNRSeg improvement, higher overall SNR improvement and higher PESQ values in objective evaluation. Another collection of spectrograms for the proposed method with other methods for speech signals corrupted with babble noise is shown in Fig. 13. This figure also attests that our proposed method has better performance in terms of harmonics' preservation and noise removal in presence of babble noise.

Formal listening tests are also conducted, where ten listeners were allowed and arranged to perceptually evaluate the enhanced speech signals. A total of thirty sentences corrupted in all background noises (airport, babble, car, exhibition, street, station and train) at SNR levels of −20 to 10 dB were processed and presented to all the listeners for evaluation. Following [16] and [23], We use SIG, BAK and OVL scales on a range of 1 to 5. The detail of these scales and procedure of this listening test is discussed in [16]. More details on this testing methodology of listening test can be obtained from [24].

The mean scores of SIG, BAK, and OVRL scales for UTM, GA and PGA, evaluated in the presence of airport noise at an SNR of 10 dB are shown in Tables 3, 4, and 5. For the three methods examined using babble noise-corrupted speech at an SNR of 10 dB, the mean scores of SIG, BAK, and OVRL scales are given in Tables 6, 7, and

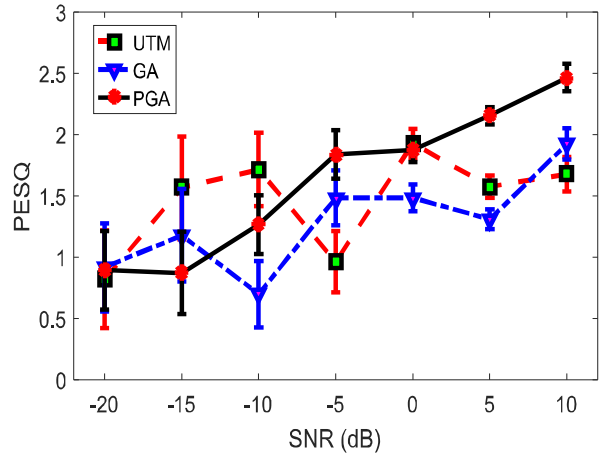

Figure 11: PESQ for different methods in babble noise.

Table 3: Mean scores of SIG scale for different methods in presence of airport noise at 10 dB

| Listener | UTM | GA  | PGA |
|----------|-----|-----|-----|
| 1        | 4.0 | 4.0 | 4.0 |
| 2        | 3.9 | 4.3 | 4.7 |
| 3        | 4.0 | 4.1 | 4.2 |
| 4        | 4.2 | 4.4 | 4.5 |
| 5        | 3.8 | 4.2 | 4.0 |
| 6        | 3.6 | 4.9 | 4.9 |
| 7        | 3.8 | 4.4 | 4.5 |
| 8        | 3.7 | 4.5 | 4.4 |
| 9        | 3.9 | 4.6 | 4.8 |
| 10       | 3.9 | 3.7 | 4.0 |

8. The mean scores in the presence of both airport and babble noises demonstrate that the lower signal distortion (i.e., higher SIG scores), the lower noise distortion (i.e., higher BAK scores) and the overall better performance (i.e., higher OVL scores) are obtained with PGA relative to those obtained by UTM and GA in most of the conditions, which attests the efficacy of the proposed method in producing better enhanced speech in comparison to other methods.

## 4. Conclusions

An improved geometric approach to spectral subtraction to solve the problem of speech enhancement has been presented in this paper. We incorporated a confidence parameter of noise estimation in the gain function of the proposed method to subtract the proper amount of noise from the magnitude of the noisy speech spectrum. Simulation results show that the proposed method yields consistently better results in terms of higher segmental SNR improvement, higher overall SNR improvement and higher PESQ values than the existing geometric approach and universal thresholding method.

[1] Y. Hu, P. C. Loizou, A generalized subspace approach for enhancing speech corrupted by colored noise, IEEE Transactions on Speech and Audio Processing 11 (4) (2003) 334–341.
[2] F. Jabloun, B. Champagne, Incorporating the human hearing properties in the signal subspace approach for speech enhancement, IEEE Trans. Speech Audio Process. 11 (2003) 700–708.

Table 4: Mean scores of BAK scale for different methods in presence of airport noise at 10 dB

| Listener | UTM | GA | PGA |
|---|---|---|---|
| 1 | 4.5 | 4.9 | 5.0 |
| 2 | 4.2 | 4.6 | 4.7 |
| 3 | 4.4 | 4.8 | 4.9 |
| 4 | 4.7 | 4.7 | 4.8 |
| 5 | 4.8 | 4.6 | 4.7 |
| 6 | 4.6 | 4.5 | 4.9 |
| 7 | 3.9 | 4.8 | 4.4 |
| 8 | 4.6 | 4.5 | 4.6 |
| 9 | 3.8 | 4.4 | 4.5 |
| 10 | 4.5 | 4.7 | 4.8 |

Table 5: Mean scores of OVL scale for different methods in presence of airport noise at 10 dB

| Listener | UTM | GA | PGA |
|---|---|---|---|
| 1 | 4.1 | 4.0 | 4.6 |
| 2 | 3.7 | 3.8 | 4.3 |
| 3 | 4.3 | 4.1 | 4.9 |
| 4 | 4.2 | 4.2 | 4.6 |
| 5 | 4.1 | 3.9 | 4.3 |
| 6 | 4.1 | 4.6 | 4.9 |
| 7 | 4.3 | 4.8 | 4.8 |
| 8 | 4.2 | 4.1 | 4.6 |
| 9 | 4.7 | 4.5 | 4.5 |
| 10 | 4.8 | 4.6 | 4.9 |

Table 6: Mean scores of SIG scale for different methods in presence of babble noise at 10 dB

| Listener | UTM | GA | PGA |
|---|---|---|---|
| 1 | 4.0 | 4.0 | 4.6 |
| 2 | 3.7 | 3.9 | 4.3 |
| 3 | 4.2 | 4.0 | 4.9 |
| 4 | 4.5 | 4.2 | 4.4 |
| 5 | 4.0 | 4.8 | 4.2 |
| 6 | 3.9 | 4.6 | 3.9 |
| 7 | 4.2 | 4.8 | 4.8 |
| 8 | 4.1 | 4.6 | 4.4 |
| 9 | 3.7 | 4.3 | 4.5 |
| 10 | 3.9 | 4.3 | 4.7 |

Table 7: Mean scores of BAK scale for different methods in presence of babble noise at 10 dB

| Listener | UTM | GA | PGA |
|---|---|---|---|
| 1 | 4.0 | 4.5 | 4.9 |
| 2 | 4.7 | 4.9 | 4.3 |
| 3 | 4.2 | 4.4 | 4.8 |
| 4 | 4.2 | 4.7 | 4.7 |
| 5 | 4.3 | 4.8 | 4.6 |
| 6 | 4.1 | 4.6 | 4.9 |
| 7 | 4.2 | 4.9 | 4.8 |
| 8 | 4.2 | 4.6 | 4.4 |
| 9 | 4.3 | 4.9 | 4.5 |
| 10 | 4.5 | 4.8 | 4.8 |

Table 8: Mean scores of OVL scale for different methods in presence of babble noise at 10 dB

| Listener | UTM | GA | PGA |
|---|---|---|---|
| 1 | 4.1 | 4.0 | 4.6 |
| 2 | 3.7 | 3.8 | 4.3 |
| 3 | 4.3 | 4.1 | 4.9 |
| 4 | 4.2 | 4.2 | 4.6 |
| 5 | 4.1 | 4.9 | 4.8 |
| 6 | 3.9 | 4.6 | 4.9 |
| 7 | 4.3 | 3.8 | 4.8 |
| 8 | 4.2 | 4.1 | 4.6 |
| 9 | 4.7 | 4.5 | 4.5 |
| 10 | 3.9 | 4.8 | 4.9 |

[3] K. S. N. You, C. H., S. Rahardja, An invertible frequency eigen domain transformation for masking-based subspace speech enhancement, IEEE Signal Process. Lett. 12 (2005) 461–464.
[4] S. Boll, Suppression of acoustic noise in speech using spectral subtraction, IEEE Transactions on acoustics, speech, and signal processing 27 (2) (1979) 113–120.
[5] K. Yamashita, T. Shimamura, Nonstationary noise estimation using low-frequency regions for spectral subtraction, IEEE Signal processing letters 12 (6) (2005) 465–468.
[6] Y. Lu, P. C. Loizou, A geometric approach to spectral subtraction, Speech communication 50 (6) (2008) 453–466.
[7] M. T. Islam, C. Shahnaz, S. Fattah, Speech enhancement based on a modified spectral subtraction method, in: 2014 IEEE 57th International Midwest Symposium on Circuits and Systems (MWSCAS), IEEE, 2014, pp. 1085–1088.
[8] M. T. Islam, A. B. Hussain, K. T. Shahid, U. Saha, C. Shahnaz, Speech enhancement based on noise compensated magnitude spectrum, in: Informatics, Electronics Vision (ICIEV), 2014 International Conference on, IEEE, 2014, pp. 1–5.
[9] Y. Ephraim, D. Malah, Speech enhancement using a minimum mean-square error log-spectral amplitude estimator, IEEE Transactions on Acoustics, Speech, and Signal Processing 33 (2) (1985) 443–445.
[10] B. Chen, P. C. Loizou, A laplacian-based (mmse) estimator for speech enhancement, Speech Commun. 49 (2007) 134–143.
[11] N. Wiener, Extrapolation, interpolation, and smoothing of stationary time series, Vol. 2, MIT press Cambridge, 1949.
[12] N. Ma, M. Bouchard, R. A. Goubran, Speech enhancement using a masking threshold constrained kalman filter and its heuristic implementations, IEEE Transactions on Audio, Speech, and Language Processing 14 (1) (2006) 19–32.
[13] D. L. Donoho, De-noising by soft-thresholding, IEEE transactions on information theory 41 (3) (1995) 613–627.
[14] M. Bahoura, J. Rouat, Wavelet speech enhancement based on the teager energy operator, IEEE Signal Process. Lett. 8 (2001) 10–12.
[15] Y. Ghanbari, M. Mollaei, A new approach for speech enhancement based on the adaptive thresholding of the wavelet packets, Speech Commun. 48 (2006) 927–940.
[16] M. T. Islam, C. Shahnaz, W.-P. Zhu, M. O. Ahmad, Speech enhancement based on student modeling of teager energy operated perceptual wavelet packet coefficients and a custom thresholding function, IEEE/ACM Transactions on Audio, Speech, and Language Processing 23 (11) (2015) 1800–1811.
[17] K. Paliwal, Estimation of noise variance from the noisy ar signal and its application in speech enhancement, IEEE Transactions on Acoustics, Speech, and Signal Processing 36 (2) (1988) 292–294.
[18] R. Martin, Spectral subtraction based on minimum statistics, power 6 (1994) 8.
[19] I. Cohen, Noise spectrum estimation in adverse environments: Improved minima controlled recursive averaging, IEEE Transactions on speech and audio processing 11 (5) (2003) 466–475.
[20] D. O'shaughnessy, Speech communication: human and machine, Universities press, 1987.
[21] Y. Hu, P. C. Loizou, Evaluation of objective quality measures for speech enhancement, IEEE Transactions on audio, speech, and language processing 16 (1) (2008) 229–238.
[22] T. Gerkmann, R. C. Hendriks, Unbiased mmse-based noise power estimation with low complexity and low tracking delay, IEEE Transactions on Audio, Speech, and Language Processing 20 (4) (2012) 1383–1393.
[23] Y. Hu, P. Loizou, Subjective comparison and evaluation of speech enhancement algorithms, Speech Commun. 49 (2007) 588–601.
[24] ITU, P835 IT: subjective test methodology for evaluating speech communication systems that include noise suppression algorithms., ITU-T Recommendation (ITU, Geneva) (2003) 835.

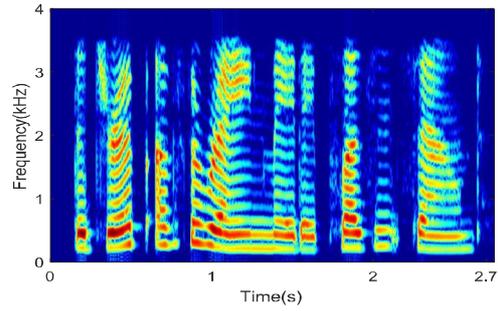

(a)

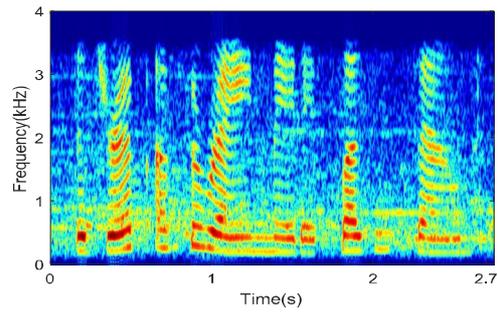

(b)

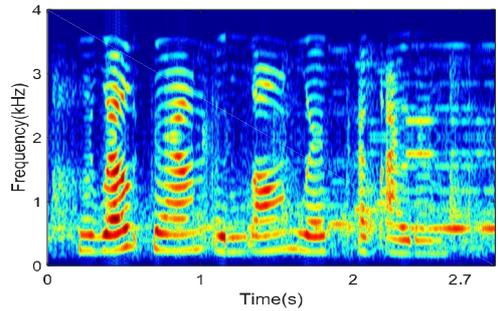

(c)

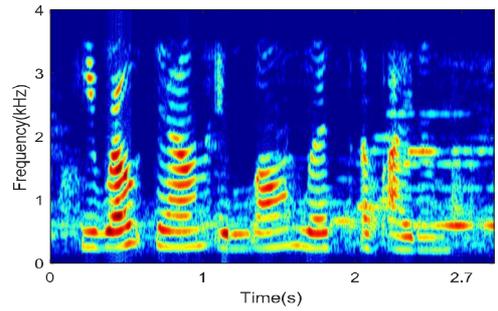

(d)

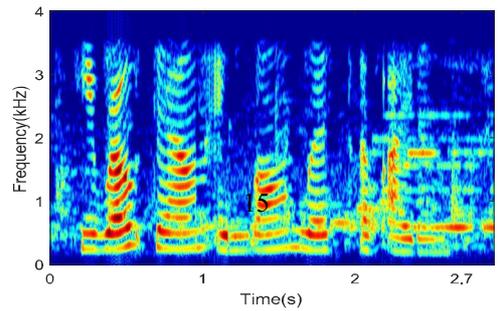

(e)

Figure 12: Spectrograms of (a) clean speech (b) noisy speech with 10 dB airport noise; spectrograms of enhanced speeches from (c) UTM (d) GA (e) PGA.

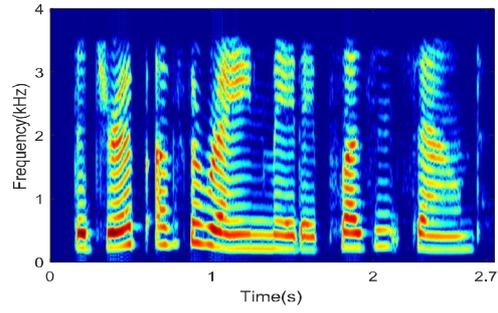

(a)

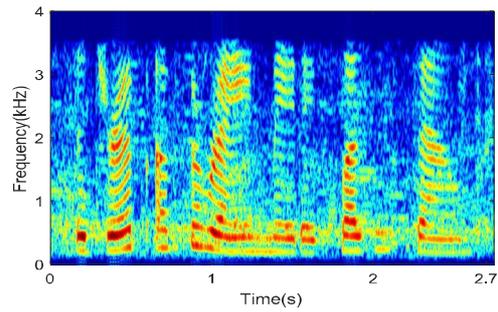

(b)

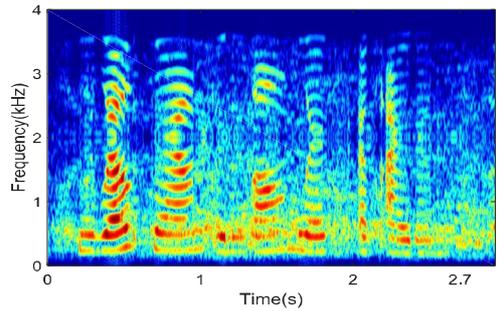

(c)

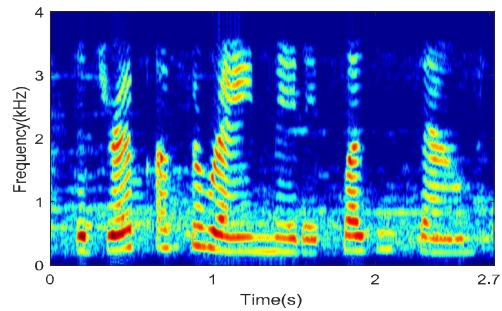

(d)

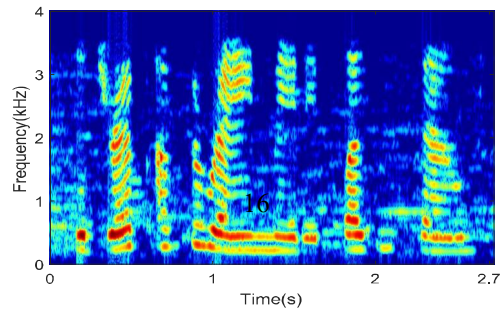

(e)

Figure 13: Spectrograms of (a) clean speech (b) noisy speech with 10 dB babble noise; spectrograms of enhanced speeches from (c) UTM (d) GA (e) PGA.

16